\documentclass[manuscript]{acmart}

\usepackage{multirow}
\usepackage{array}
\usepackage{graphicx}
\usepackage{subcaption}
\usepackage[inline]{enumitem}
\usepackage[utf8]{inputenc}

\setcopyright{acmcopyright}
\copyrightyear{2018}
\acmYear{2018}
\acmDOI{10.1145/1122445.1122456}

\acmConference[Woodstock '18]{Woodstock '18: ACM Symposium on Neural
  Gaze Detection}{June 03--05, 2018}{Woodstock, NY}
\acmBooktitle{Woodstock '18: ACM Symposium on Neural Gaze Detection,
  June 03--05, 2018, Woodstock, NY}
\acmPrice{15.00}
\acmISBN{978-1-4503-XXXX-X/18/06}

\begin{document}

\title{Recommending Burgers based on Pizza Preferences: Addressing Data Sparsity with a Product of Experts}

\author{Martin Milenkoski}
\email{martin.milenkoski@epfl.ch}
\authornote{Work done while at Swisscom.}
\author{Diego Antognini}
\email{diego.antognini@epfl.ch}
\affiliation{%
  \institution{
École Polytechnique Fédérale de Lausanne}
  \city{Lausanne}
  \country{Switzerland}
}

\author{Claudiu Musat}
\email{claudiu.musat@swisscom.com}
\affiliation{%
  \institution{Swisscom}
  \city{Lausanne}
  \country{Switzerland}
}

\begin{abstract}
In this paper, we describe a method to tackle data sparsity and create recommendations in domains with limited knowledge about user preferences. We expand the variational autoencoder collaborative filtering from a single-domain to a multi-domain setting. The intuition is that user-item interactions in a source domain can augment the recommendation quality in a target domain. The intuition can be taken to its extreme, where, in a cross-domain setup, the user history in a source domain is enough to generate high-quality recommendations in a target one. We thus create a Product-of-Experts (POE) architecture for recommendations that jointly models user-item interactions across multiple domains. The method is resilient to missing data for one or more of the domains, which is a situation often found in real life. 
  We present results on two widely-used datasets - Amazon and Yelp, which support the claim that holistic user preference knowledge leads to better recommendations. Surprisingly, we find that in some cases, a POE recommender that does not access the target domain user representation can surpass a strong VAE recommender baseline trained on the target domain.
\end{abstract}

\keywords{Single-Domain Recommendation, Cross-Domain Recommendation, Variational Autoencoder}

\maketitle

\section{Introduction}

Recommender systems are ubiquitous. Due to their prevalence, the user-item interactions are becoming less sparse~and collaborative approaches gain the upper hand on content-based methods. This has led to a plethora of neural collaborative filtering methods \cite{antognini2020interacting,he2017neural,liang2018variational}. 
However, they suffer from the cold-start problem. Traditionally, it can be alleviated with smarter priors and better defaults. For instance, content-based methods can be integrated as features in complex systems that merge different knowledge sources \cite{Zhou_2020}. Other examples are preference elicitation~\cite{Chajewska00makingrational,Chajewska98utilityelicitation} or critiquing \cite{unitcritiquing,reilly2005explaining}, where users express their preferences by interacting with the items' attributes. A major drawback is the assumption of a fixed set of known attributes, especially for domains with expressive features such as hotels or products~\cite{ni2019justifying,hotelrec}.

There is, however, no escaping the fact that with no knowledge about the users, personalization is impossible. As personalization is what generally makes recommenders useful, we inquire what are the additional information sources that enable it even in the absence of an in-domain interaction history. In this line of thought, known as Cross-domain recommendations \cite{10.1145/1401890.1401969,Zhao10multi-domaincollaborative}, we assume consistency between the behaviors of users across different domains. The consistency rests on the belief that observable behaviors in one domain depend partially on some core values and~beliefs.

The idea of cross-domain recommendations is not new \cite{hu2013personalized, loni2014cross, hu2018conet, yuan2019darec}. 
The main issue with most of these approaches~is that they can only be applied in a fully-supervised setting. This means that the model needs to be trained on users with existing interactions on all the considered domains. This is not aligned with one very common use case, when a user has an extensive history in one domain (e.g. buying pizza) but no history in another (e.g. buying~burgers).

To fully use existing user-item interactions, we believe a cross-domain recommender system needs to handle cases where user interactions are lacking for one more multiple domains both during training and during inference. To achieve this, we propose a Product of Experts (POE) method, inspired by previous work on multimodal variational autoencoders \cite{wu2018multimodal}.  
By assuming conditional independence between the domains, given the common latent representation, we can ignore the missing domains when calculating the joint latent distribution. We model the joint distribution as a product of experts of the latent distributions of the individual domains
. Based on prior work on the usage of variational autoencoders for collaborative filtering \cite{liang2018variational}, we assume that each latent distribution is Gaussian. Also, these assumptions allow us to calculate analytically the joint distribution because the product of Gaussian distributions is itself a Gaussian~\cite{cao2014generalized}. 


We evaluate the performance of the POE model on two different datasets containing Amazon and Yelp reviews. Both datasets contain multiple domains, for instance books and clothing for Amazon and restaurants and shopping for Yelp. We experiment using combinations of two domains at a time, with the combinations ranging from very similar domains - like Books and Kindle in Amazon, to very dissimilar ones, like Books and Clothing.
We measure the performance in terms of Recall and NDCG. We do so in two different but related settings. First, we focus on single-domain recommendation accuracy, by training the recommender simultaneously on a pair of domains. The second setting is the pure cross-domain recommendation, where we make recommendations in a target domain for a user without knowing their history in that domain. We train the model with user interactions from both a source and target domains but recommend items for the target domain, only based on the user's interactions in the source one.

Our experiments demonstrate that in a single domain recommendation setup, our method outperforms single-domain VAEs on the majority of cases. In cross-domain recommendations, we show that even completely ignoring the knowledge about a user on the target domain, our model generates recommendations significantly better than the unpersonalized baseline, getting close and even surpassing the recommendation quality obtained with an in-domain~VAE.

\section{Related Work}
\label{related-work}

There has been a large amount of research dedicated to the usage of Variational Autoencoders in the setting of single-domain recommender systems \cite{liang2018variational, li2017collaborative, chen2018collective, he2019collaborative}. 
Our work is a direct extension of \cite{liang2018variational} in the setting of cross-domain recommendation. To handle data sparsity, several studies focus on extending the VAE approach in the hybrid setting, incorporating item content as a side information in the recommendation process \cite{li2017collaborative, chen2018collective, he2019collaborative, 10.1145/3383313.3412256}. Similarly, our work incorporates side information as well. However, we leverage user interactions on other domains instead of item  features which might not always be available. Additionally, our model supports weakly-supervised learning and does not require users to have interactions in all domains. Once trained, the model can be used as a single-domain recommender system using only data from the domain of interest, eliminating the need for side information in the recommendation process. 


More recent work has been focused on learning the joint distribution $p(x_1, x_2)$ explicitly using a joint inference network $q(z | x_1, x_2)$ with two additional inference networks $q(z | x_1)$ and $q(z | x_2)$ to handle missing data during inference~\cite{suzuki2016joint,vedantam2017generative}. 
However, the main drawback of such approaches is that they scale poorly with the number of modalities and are intractable in a general setting. A significant step forward has been made with the multimodal variational autoencoder (MVAE) \cite{wu2018multimodal}. MVAE approximates the joint posterior distribution as Product of Experts over the marginal posteriors. This enables cross-modal generation at inference time without the need of additional inference networks. \cite{shi2019variational} proposed a mixture of experts model (MMVAE). Its main drawback is that it can only be trained when data is available for all modalities. \cite{sutter2020multimodal} proposes a novel objective function that utilizes Jenson-Shannon divergence for multiple distributions and a dynamic prior.
In this work, we build upon MVAE and adapt it for single and cross-domain~recommendation.

The sets of users (or items) in cross-domain recommendation systems might be disjoint, overlap partially, or be the same. The work in this paper focuses on the case of partially overlapping users and items. Early work on this use case has been focused on adopting matrix factorization and transferring shared knowledge based on overlapping users or items \cite{pan2010transfer, singh2008relational,hu2018conet}. 
The closest work to ours is \cite{nguyen2018domain}, where the authors build a cross-domain recommendation framework consisting of domain-specific VAEs for encoding the user interaction vectors and generative adversarial networks for generating user interaction vectors. 
In our work, we approximate the true joint latent distribution with a product of experts. 
This allows us to generate recommendations from and to each domain with both partial and full input. Additionally, we propose a cross-domain recommendation model based solely on VAEs without the need for additional networks. Finally, no additional effort in the modeling is needed to extend the solution to more than two~domains. 

\section{Method}
\label{method}
Before proceeding, we define the following notation used throughout this paper:
\begin{itemize}
    \item $U$, $I$, and $D$ are the number of users, items, and domains. $U_d$ and $I_d$ are the user and item subsets in domain $d$.
    \item $\mathbf{x_u^d}$: A binary vector with length $I_d$. This is the implicit feedback vector for user $u$ on domain $d$.
    \item $N_u^d$: Total number of interactions of user $u$ with domain $d$. $N_u^d = \sum_i x_{u, i}^d$
\end{itemize}

\subsection{Variational Autoencoders for Collaborative Filtering}
\label{vae}

A variational autoencoder (VAE) is a generative model that has been successfully applied in the setting of single-domain collaborative filtering task \cite{liang2018variational, chen2018collective, li2017collaborative}. Prior work \cite{liang2018variational} focuses on the application of VAE on single-domain recommendation. Here, we present their work contextualized in the setting of cross-domain recommendation. We explicitly denote the single domain of interest as $d$ and the user feedback vector as $x_u^d$. For each user $u$, the model samples the latent representation $\mathbf{z_u^d}$ from a standard Gaussian prior. Then, a non-linear function $f_{\theta_d}(\cdot)$ is applied on $\mathbf{z_u^d}$ and normalized via softmax to produce a probability distribution $\pi^d(z_u^d)$. The function $f_{\theta_d}(\cdot)$ is a neural network with parameters $\theta_d$. The feedback vector $x_u^d$ is assumed to be drawn from a multinomial distribution with probability~$\pi^d(z_u^d)$:\begin{equation}
\small
z_u^d \sim \mathcal{N}(0, I_k),
    \quad\quad
    \pi^d(z_u^d) \propto exp\{f_{\theta_d}(z_u^d)\}, 
    \quad \quad
    x_u^d \sim Mult(N_{u}^d, \pi^d(z_u^d))
    \quad\quad
    \log p_{\theta_d}(x_u^d|z_u^d) = \sum_{i=1}^{I_d} x_{u, i}^d \log \pi_i^d(z_u^d)
\label{generative-model}
\end{equation}
%
To learn the generative model
we need to 
approximate the intractable posterior distribution $p(z_u^d | x_u^d)$. \cite{liang2018variational} approximates the posterior distribution using a variational distribution $q(z_u^d)$ learned with an inference model defined as follows:
\begin{equation}
\small
    g_{\phi_d}(x_u^d) = [\mu_{\phi_d}(x_u^d), \sigma_{\phi_d}(x_u^d)] \in R^{2k}, \quad q_{\phi_d}(z_u^d|x_u^d) = \mathcal{N}(\mu_{\phi_d}(x_u^d), diag\{ \sigma_{\phi_d}^2(x_u^d)\})
\label{latent-distribution}
\end{equation} In this way, using the input $x_u^d$, the inference model outputs the parameters of the variational distribution $q_{\phi_d}(z_u^d|x_u^d)$, which approximates the posterior distribution $p(z_u^d | x_u^d)$. 
Finally, the objective function is the evidence lower bound~(ELBO):
\begin{equation}
    \small
    \mathcal{L}(x_u^d;\theta_d,\phi_d) = \mathbb{E}_{q_{\phi_d}(z_u^d|x_u^d)}[\log{p_{\theta_d}(x_u^d|z_u^d)}] - \beta \cdot KL(q_{\phi_d}(z_u^d|x_u^d)||p(z_u^d))
\label{elbo-equation}
\end{equation}
The first term can be interpreted as negative reconstruction error and the second term can be interpreted as a regularization term, which is the KL divergence between the variational distribution $q_{\phi_d}(z_u^d|x_u^d)$ and the prior $p(z_u^d)$. The parameter $\beta$ controls the strength of the regularization and is tuned using a KL annealing strategy \cite{bowman2015generating}.

\subsection{Variational Autoencoders for Cross-Domain Recommendation}
\label{vae-cross}

A simple extension of the VAE model in the setting of cross-domain recommendation is treating the set of domains $\{1,...,D\}$ as a single domain $\widetilde{d}$. In this case, the input for user $u$ would be a concatenated vector of the individual domain vectors $x_u^{\widetilde{d}} = (x_i^1, ..., x_i^D)$. The dimensionality of the input vector would be $\sum_d I_d$. By setting $d=\widetilde{d}$ in Equation~\ref{generative-model}-\ref{elbo-equation}, we can use the same procedure described in Section~\ref{vae} to train a single-domain recommender on the merged domain~$\widetilde{d}$. 

The main issue with this approach is that it can only be applied in a fully-supervised setting. This means that the model needs to be trained on users with feedback in all domains $D$. Additionally, since the model is never trained with individual domains, it will not be able to make predictions for users with feedback on only one domain. This fact severely limits the applicability of this approach. We remediate this problem in the next section. 

\subsection{Product of Experts Variational Autoencoder}

In real-life applications, users may have no history on a given domain. For example, on an e-commerce website users might purchase many items from a single category, but they might have never purchased an item from other categories. To make the most effective use of the available data, a cross-domain recommender system needs to handle missing user feedback on some domains both during training and during inference. To tackle this problem, we propose a novel model for cross-domain recommendation inspired by previous work on multimodal variational autoencoders \cite{wu2018multimodal, sutter2020multimodal, shi2019variational}.  

\subsubsection{Model}
The proposed model is based on a product of experts variational autoencoder (POE). 
It is an extension of the collaborative variational autoencoder in the setting of weakly-supervised cross-domain recommendation. In this setting, we work with a set of user feedback vectors $\{x_u^d\}$, where $d \in \{1,...,D\}$. As in the multimodal setting \cite{wu2018multimodal}, we assume that the $D$ domain feedback vectors $x_u^d$ are conditionally independent given the common latent variable $z_u$. In other words, we assume a generative model of the form
$
p_\theta(x_u^1, ..., x_u^D, z_u) = p(z_u)p_{\theta_1}(x_u^1|z_u)...p_{\theta_D}(x_u^D|z_u)
$
where $\theta_d$ are the parameters of the generative model (decoder) associated with domain~$d$. With this factorization, we can ignore missing domains when calculating the marginal likelihood. If a user $u$ has no feedback for domain $d$, we can omit the term $p_{\theta_d}(x_u^d|z_u)$ from the generative model and still use the model with the known domains for user $u$. 

As in the single-domain setting, for each user u, the model samples a latent representation $z_u$ from a standard Gaussian prior. Then, we define $D$ non-linear functions $f_{\theta_d}(\cdot)$ to produce $D$ propabibility distributions $\pi^d(z_u)$. The function $f_{\theta_d}(\cdot)$ is a domain-specific decoder with parameters $\theta_d$. The feedback vector $x_u^d$ is assumed to be drawn from a multinomial distribution with probability $\pi^d(z_u)$:
\begin{equation}
\small
z_u \sim \mathcal{N}(0, I_k),
\quad
    \pi^d(z_u) \propto exp\{f_{\theta_d}(z_u)\}, 
    \quad 
    x_u^d \sim Mult(N_{u}^d, \pi^d(z_u))
\end{equation}

\subsubsection{Variational inference}
The main issue for training with missing domains is specifying the $2^D$ inference networks $q(z_u|X_u)$ for each subset of domains $X_u \subseteq \{x_u^1, ..., x_u^D\}$. Prior work on multimodal learning \cite{wu2018multimodal} has shown that under the assumption of conditional independence, the joint posterior distribution can be approximated by a product of~experts:\begin{equation}
\small
    q_\phi(z_u|X_u) \propto p(z_u) \prod_{x_u^d \in X_u} q_{\phi_d}(z_u^d|x_u^d) 
\end{equation}
where $p(z_u)$ is a prior expert, and $q_{\phi_d}(z_u^d|x_u^d)$ is the domain-specific inference network (encoder) for domain $d$ as defined in Equation~\ref{latent-distribution}. The prior expert is a standard Gaussian distribution. The product distribution described above is not solvable in closed form in a general case. However, when both $p(z_u)$ and $q_{\phi_d}(z_u|x_u^d)$ are Gaussian, their product is itself a Gaussian \cite{cao2014generalized} with mean $\mu$ and covariance $V$ defined as: $\mu = (\sum_d \mu_dV_d^{-1})(\sum_d V_d^{-1})^{-1}, V = (\sum_d V_d^{-1})^{-1}
$, where $\mu_d, V_d$ are the parameters of the latent distribution for domain $d$. In our case, both the prior expert and the inference networks are Gaussian distributions. For this reason, we can compute the mean and covariance of the product distribution as described above. Therefore, we can avoid specifying the $2^D$ inference networks and train the model efficiently in terms of the D inference networks $q_{\phi_d}(z_u|x_u^d)$.

\subsubsection{Training}
Let us define the set of user feedback vectors used as input $X_u = \{x_u^d |$ domain $d$ used as input for user~$u$$\}$ and the set of user feedback vectors present in the data $\widetilde{X_u} = \{x_u^d | $ domain $d$ present for user $u$$\}$. We use the following form of the ELBO objective for a given input $X_u$:
\begin{equation}
\small
  \mathcal{L}(X_u;\theta,\phi) = \mathbb{E}_{q_\phi(z_u|X_u)}[\sum_{x_u^d \in \widetilde{X_u}} \lambda_d \log{p_{\theta_d}(x_u^d|z_u)}] - \beta * KL(q_\phi(z_u|X_u)||p(z_u))  
\label{elbo-loss}
\end{equation}
In multimodal learning, prior work consider $X = \widetilde{X}$ \cite{wu2018multimodal}. However, treating them as different allows us to teach the model to generate better cross-domain recommendations when using individual domains as input. In this way, when presented with a single domain as input, the model is taught to perform well on all known domains instead of only the domain used as input. $\lambda_d$ is the weight given to domain $d$ in the loss function. 

Optimizing this function on a dataset with fully present domains for all users has an undesirable effect; since the model does not see data points with missing domains during training, it cannot do inference with individual domains. For this reason, we use a modified version of the sub-sampled objective in \cite{wu2018multimodal} to train the model for individual inputs:
\begin{equation}
\small
\mathcal{L}(\{x_u^1,...,x_u^D\};\theta,\phi) + \sum_{d=1}^D \mathcal{L}(\{x_u^d\};\theta,\phi)    
\end{equation}

\subsubsection{Inference}

We now describe how to make predictions given a trained POE model. We describe the prediction process in the setting of two domains ($D=2$). Without loss of generality, let us define $s=d$ as the source domain, and $t=d$ as the target domain. In this paper, we present two applications of the model - single-domain and cross-domain recommendation. In the single-domain recommendation setting, we recommend items from the target domain using the user feedback from the target domain. In the cross-domain recommendation setting, we recommend items from the target domain using the user feedback from the source domain. 
 
 Let us define the domain used as input as $v$, with $v=t$ in the single-domain setting, and $v=s$ in the cross-domain setting. In both cases, we pass the input $x_u^v$  through the inference model $g_{\phi_v}(x_u^v)$ to obtain the parameters $\mu_{\phi_v}$ and $\sigma_{\phi_v}$. Then, we calculate the product of experts distribution and take the mean of the distribution as the latent representation $z_u$.  Finally, we rank the items in the target domain based on the un-normalized predicted multinomial probability~$f_{\theta_t}(z_u)$. 

\section{Experiments}
\label{experiments}
%
\begin{table*}[!t] 
\footnotesize
\center
\caption{Descriptive statistics of the datasets after preprocessing. Interactions are non-zero entries. }
\begin{tabular}{>{\raggedright}p{0.02\linewidth}>{\raggedright}p{0.1\linewidth}>{\raggedright}p{0.12\linewidth}>{\raggedright}p{0.1\linewidth}>{\raggedright}p{0.1\linewidth}>{\raggedright}p{0.1\linewidth}>{\raggedright}p{0.12\linewidth}>{\raggedright\arraybackslash}p{0.12\linewidth}}
\# & Dataset & Category & Item threshold & \# of users &  \# of items &  \# of interactions &  Density  \\
\midrule
\multirow{2}*{1} & \multirow{2}*{Amazon}      & Books       &  200  & \multirow{2}*{63,711} & 29,124 & 2,041,610 & 0.11\%    \\
     & & Kindle Store       &  30   &     &   30,243   & 1,014,985& 0.05\% \\
     \midrule
     \multirow{2}*{2} & \multirow{2}*{Amazon}      & Books       &  200  & \multirow{2}*{43,242} & 29,266 & 702,081 &   0.06\%  \\
     & & Movies       &  20   &     &   33,793   & 671,961 & 0.05\%\\
     \midrule
     \multirow{2}*{3} & \multirow{2}*{Amazon}      & Books       &  200  & \multirow{2}*{42,965} & 29,354 & 602,743 & 0.05\%    \\
     & & Clothing       &  150   &     &   24,244   & 390,677 & 0.04\%\\
     \midrule
     \multirow{2}*{4} & \multirow{2}*{Yelp}      & Restaurants       &  100  & \multirow{2}*{86,566} & 7,886 & 599,587 & 0.09\%    \\
     & & Hotels       &  1   &     &   5,089   & 142,516 & 0.03\%\\
     \midrule
     \multirow{2}*{5} & \multirow{2}*{Yelp}      & Restaurants       &  10  & \multirow{2}*{138,801} & 35,361 & 1,260,613 & 0.03\%    \\
     & & Shopping       &  1   &     &  30,998   & 291,897 & 0.01\%\\
     \midrule
     \multirow{2}*{6} & \multirow{2}*{Yelp}      & Food       &  1  & \multirow{2}*{110,427} & 29,929 & 577,291 & 0.02\%    \\
     & & Shopping       &  1   &     &   30,171   & 256,196 & 0.01\%\\
     \midrule
     \multirow{2}*{7} & \multirow{2}*{Yelp}      & Burgers       &  1  & \multirow{2}*{67,088} & 5,340 & 143,799 & 0.04\%    \\
     & & Pizza       &  1   &     &   6,364   & 151,864 & 0.04\%\\
\label{data-attributes}
\end{tabular}
\end{table*}
\subsection{Datasets}
In order to evaluate the quantitative performance of our proposed POE model, we perform experiments using seven real-world publicly available datasets based on Amazon reviews \cite{ni2019justifying} and Yelp businesses.
All datasets contains user and item (partial) overlaps. This means that both users and items can be present in more than one domain.

\textbf{Amazon Reviews} dataset contains information regarding product reviews and metadata obtained from Amazon~\cite{ni2019justifying}. IWe consider each category as separate domain. We chose three of the most popular categories Books, Movies and Clothing to maximize user overlap. Additionally, we chose Kindle Store as a semantically related category with Books in order to study the effect of semantic similarity. We binarize the reviews to create a dataset with implicit feedback. Only reviews $\ge3.5$ are considered as a signal that the user liked the item. All users with less than 5 reviews are filtered out.\\
\indent \textbf{Yelp dataset} contains information about businesses, reviews and user data on Yelp \cite{yelpdataset}. We choose some of the most popular categories in order to maximize user overlap. The ratings are binarized to create a dataset with implicit feedback. Only ratings $\ge3.5$ were considered as a signal that the user liked the business. 

\subsection{Metrics}

A trained POE model on $D$ domains can be used to produce single-domain recommendation for all domains $D$. For this reason, a single model can be evaluated on the metric of interest for all domains. To present the performance of a single model on all domains we use the concept of Pareto front. Formally, let $M$ be the metric of interest and let $w_1$ and $w_2$ be $D$-dimensional vectors associated with two models $m_1$ and $m_2$. $w_{i, d}$ is the performance of model $m_i$ on domain $d$ on the metric $M$. The model $m_1$ is said to dominate the model $m_2$ if $\forall d \in \{1,...,D\} \: w_{1, d} \geq w_{2, d}$ and $\exists d \in \{1,...,D\} \: w_{1, d} > w_{2, d}$. A Pareto optimal model is a model $m^*$ which is not dominated by any other model $m$. The set of all Pareto optimal models is called a Pareto set and its visualization is called a Pareto front. As metrics, we use Recall@K and NDCG@K. 

\subsection{Experimental Details}

We run the experiments on 7 domain combinations. For each combination of categories we choose a category-specific threshold and we filter out all items with number of reviews less than that threshold. The choice of the threshold was dependent on the domain combination and was usually tuned to bring the complexity of the domains close to each other. Table \ref{data-attributes} presents the postprocessing attributes of the datasets. We use the same model architecture and hyperparameters for both the VAE and POE approach in order to obtain fair comparison. With the large combination of applications, domains, and combinations, we employed the default parameters used in \cite{liang2018variational}. 
We use 95\% of users as a train set and 5\% as a test set. In the test set, 20\% of user interactions are masked and used as ground truth to evaluate the model. The remaining 80\% of interactions are used as input. 

\subsection{Single-Domain Recommendation}
\label{single-domain-section}

\begin{figure*}[t]
\centering
\begin{subfigure}[]{0.32\textwidth}
    \centering
    \includegraphics[width=\textwidth,height=.99in]{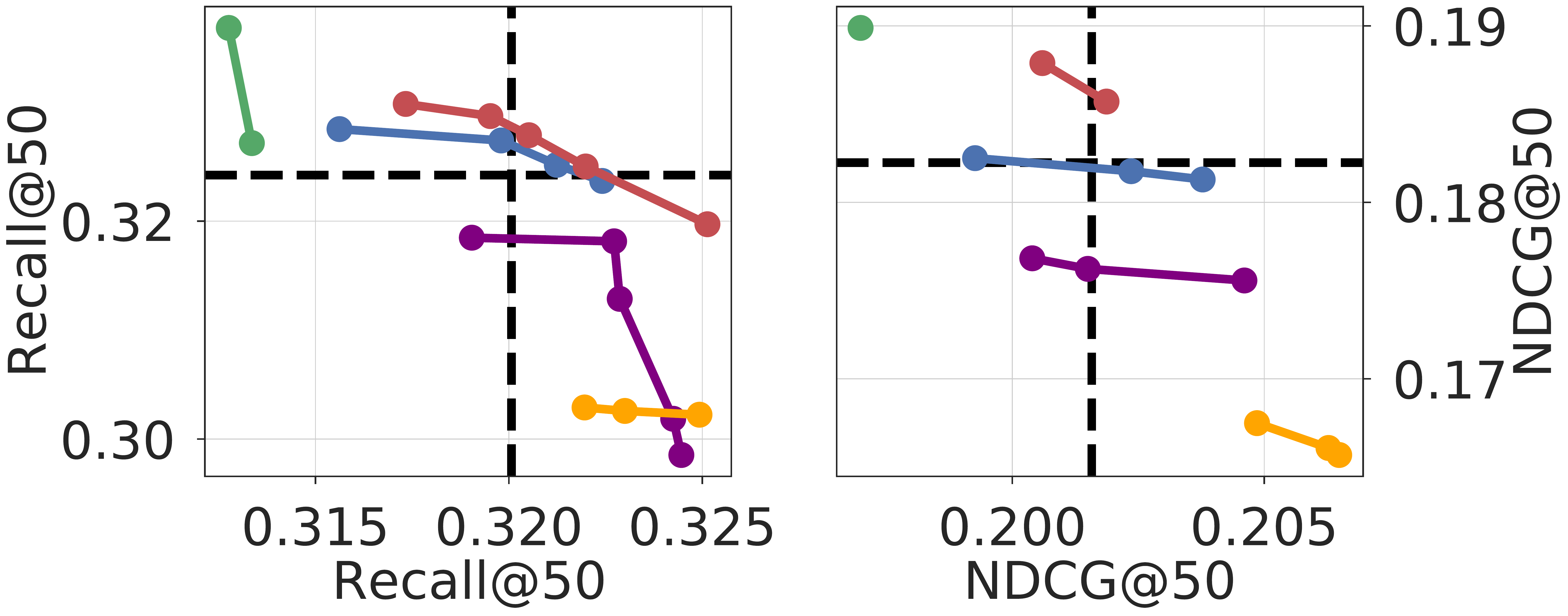}
    \caption{\textbf{Books and Kindle}.}
\end{subfigure}
~
\begin{subfigure}[]{0.32\textwidth}
    \centering
    \includegraphics[width=\textwidth,height=.99in]{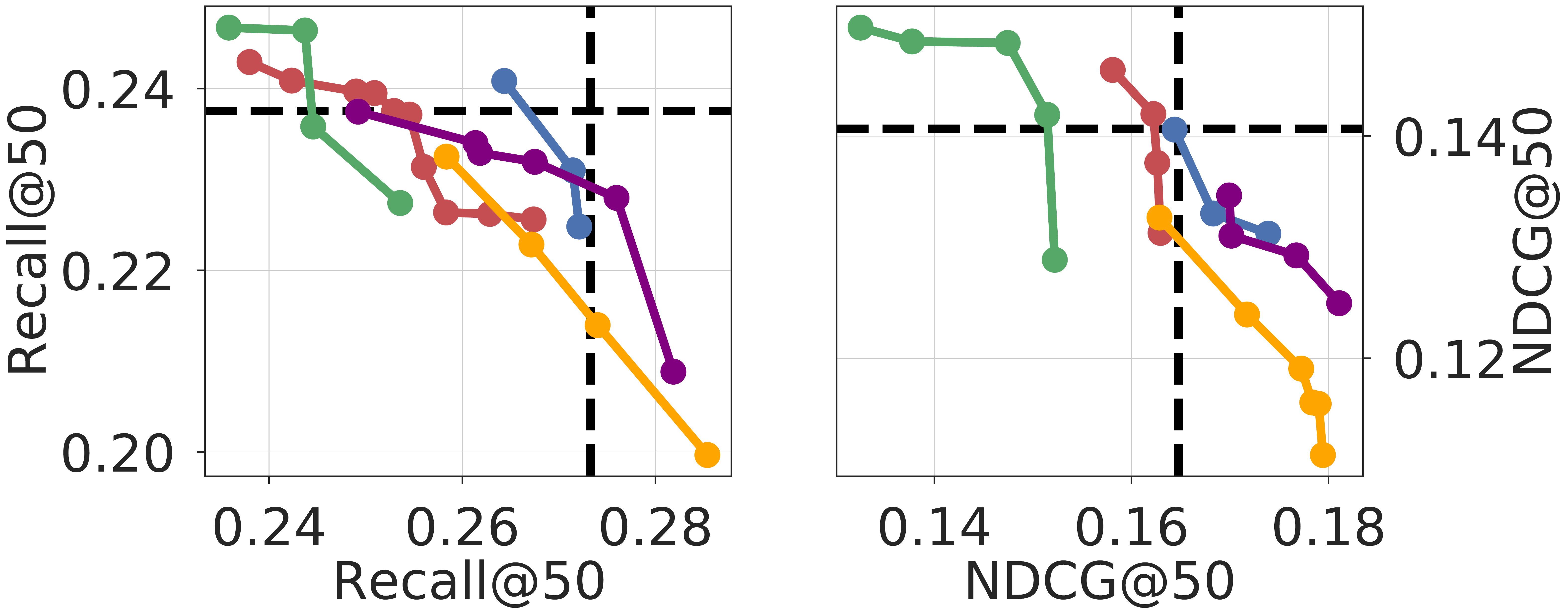}
    \caption{\textbf{Books and Movies}.}
\end{subfigure}
~
\begin{subfigure}[]{0.32\textwidth}
    \centering
    \includegraphics[width=\textwidth,height=.99in]{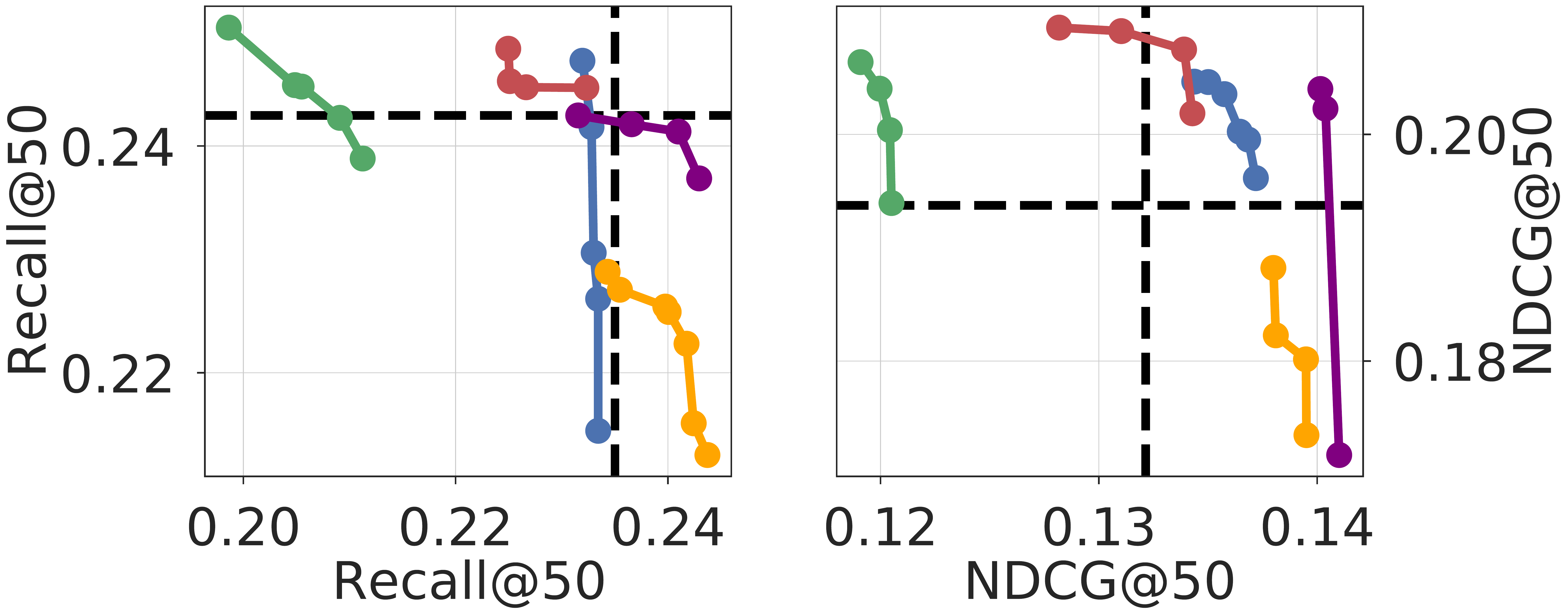}
    \caption{\textbf{Books and Clothing}.}
\end{subfigure}
\\
\begin{subfigure}[]{0.32\textwidth}
    \centering
    \includegraphics[width=\textwidth,height=.99in]{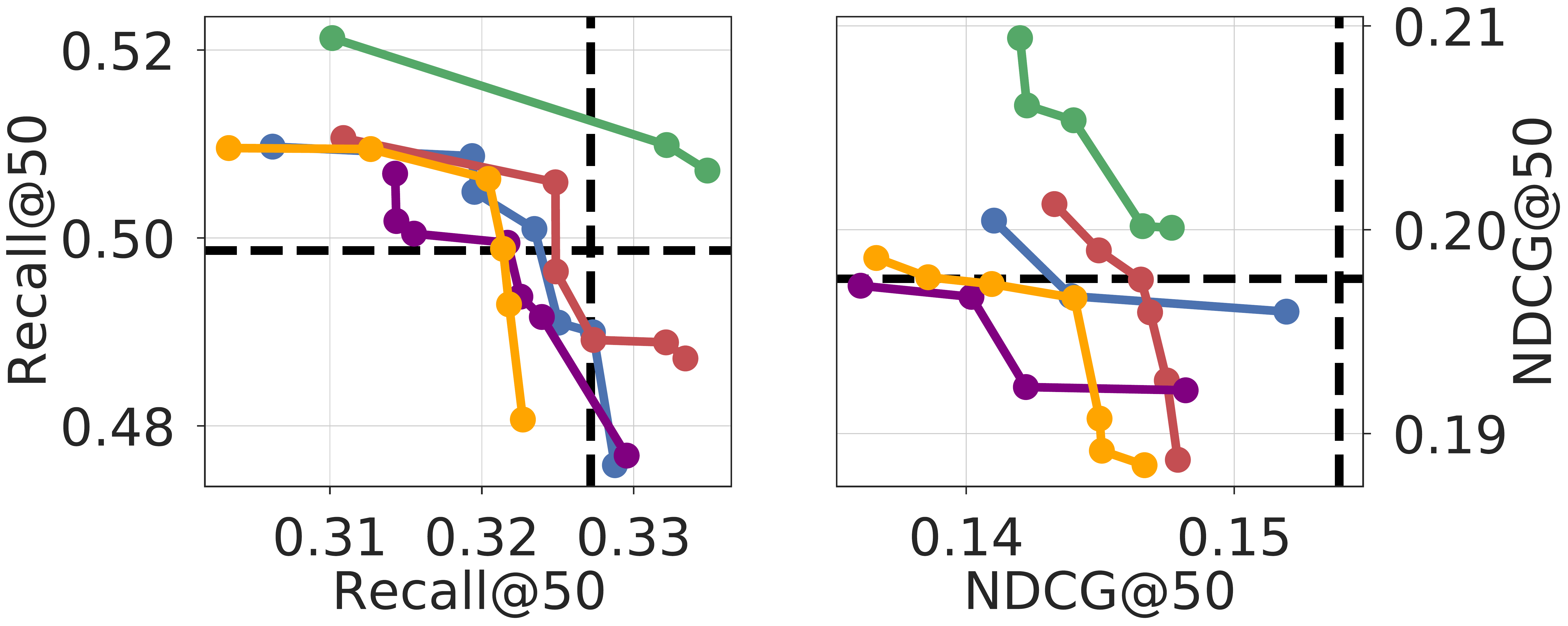}
    \caption{\textbf{Restaurants and Hotels}.}
\end{subfigure}
~
\begin{subfigure}[]{0.32\textwidth}
    \centering
    \includegraphics[width=\textwidth,height=.99in]{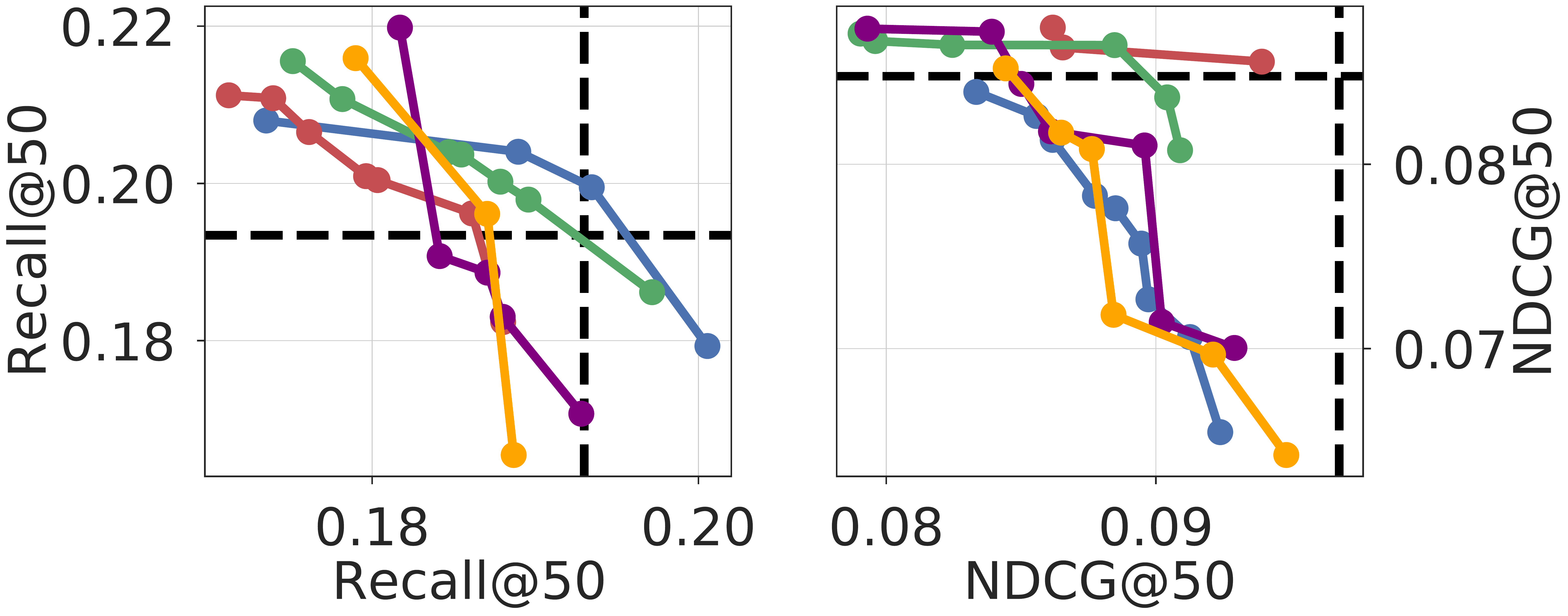}
    \caption{\textbf{Restaurants and Shopping}.}
\end{subfigure}
~
\begin{subfigure}[]{0.32\textwidth}
    \centering
    \includegraphics[width=\textwidth,height=.99in]{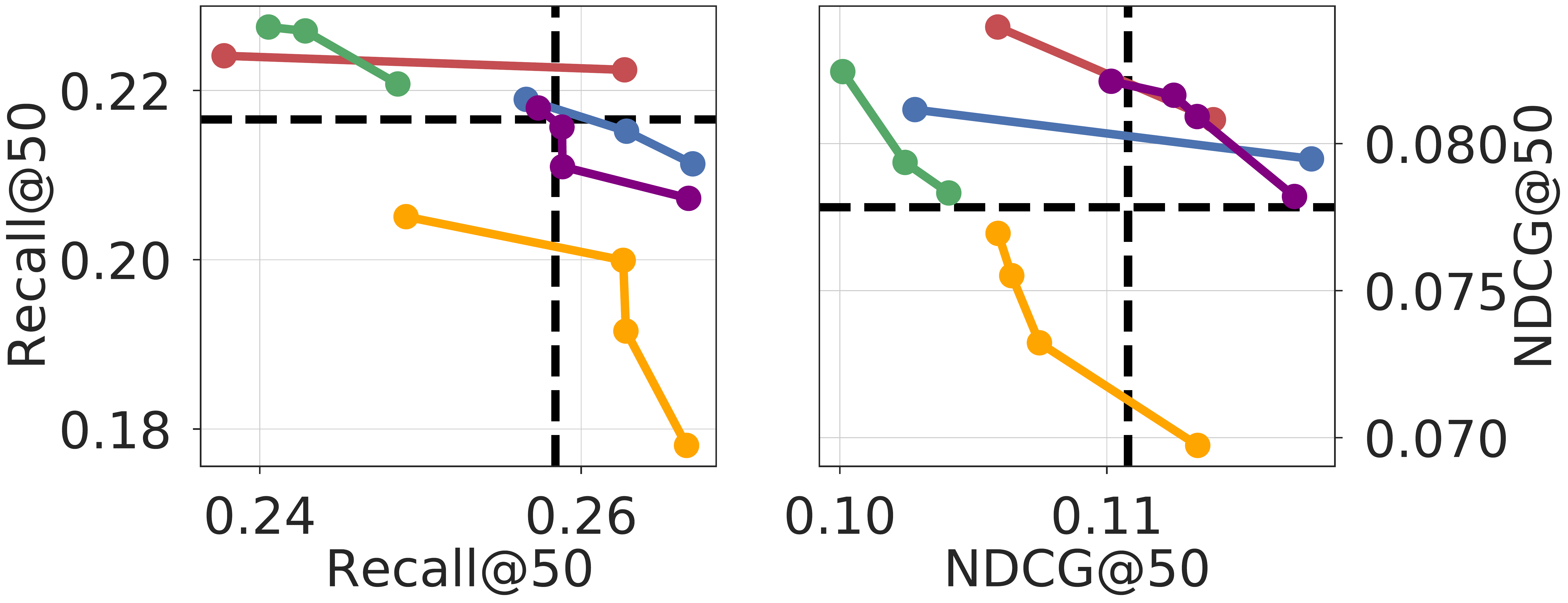}
    \caption{\textbf{Food and Shopping}.}
\end{subfigure}\\
\begin{subfigure}[]{0.32\textwidth}
    \centering
    \includegraphics[width=\textwidth,height=0.99in]{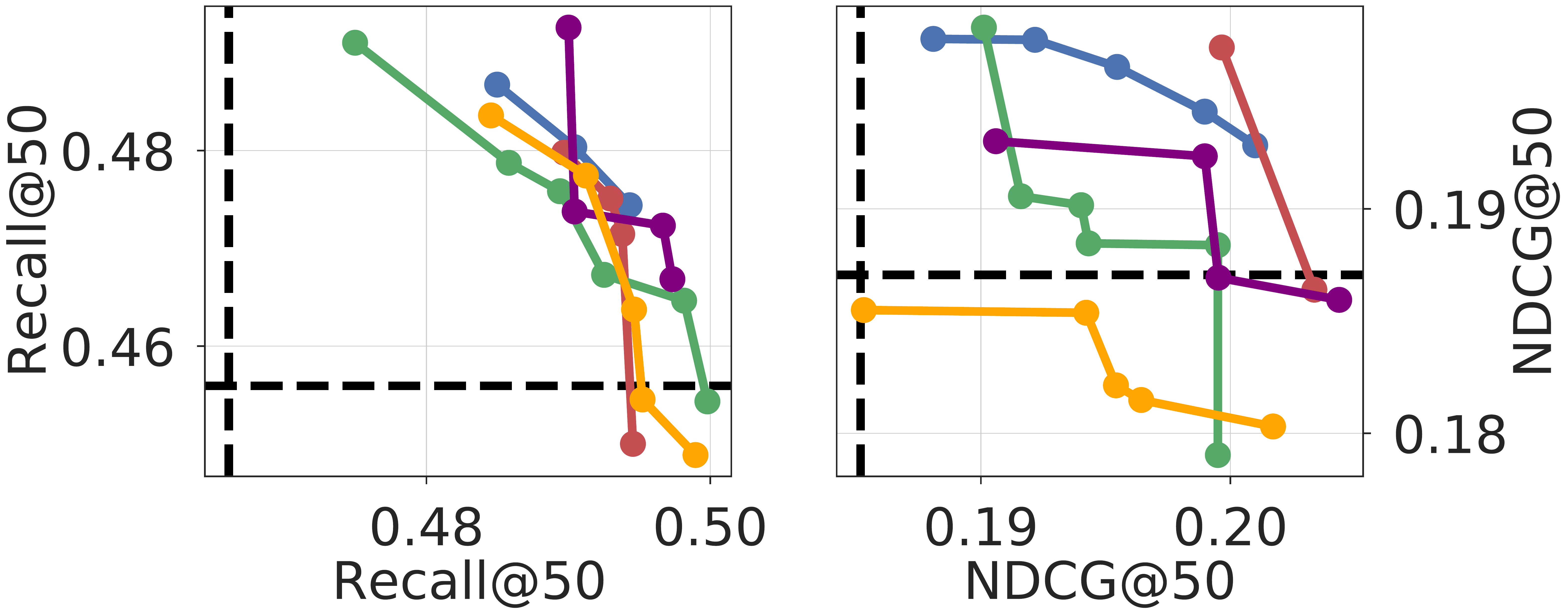}
    \caption{\textbf{Burgers and Pizza}.}
\end{subfigure}
~
\begin{subfigure}[]{0.32\textwidth}
\end{subfigure}
~
\begin{subfigure}[t]{0.32\textwidth}
    \centering
    \includegraphics[width=\textwidth]{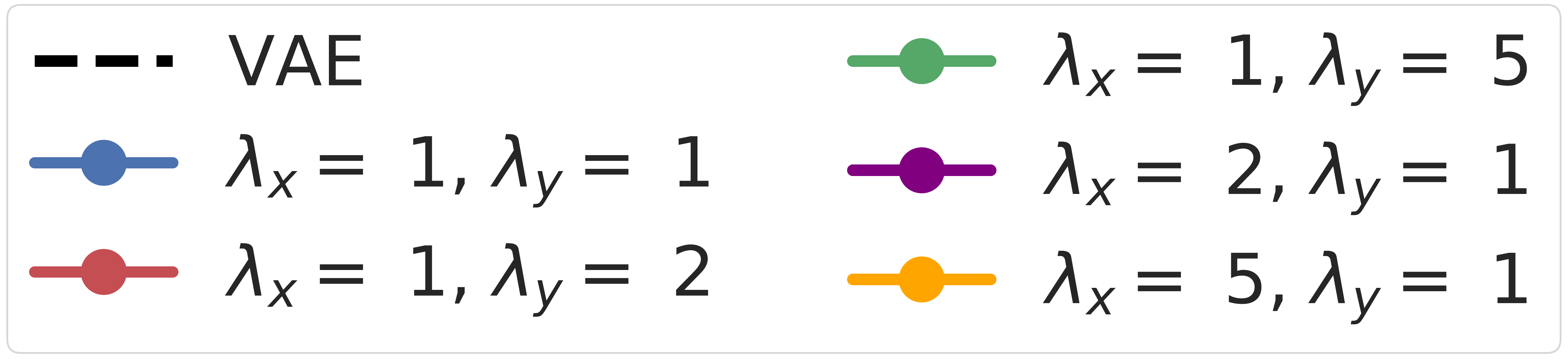}
\end{subfigure}
\caption{Pareto fronts for Recall@50 and NDCG@50 on all 7 datasets with different domain weights.}
\label{pareto-fronts}
\end{figure*}

In the single-domain setting we compare the POE model to a traditional single-domain VAE model from \cite{liang2018variational}. On Figure~\ref{pareto-fronts}, the VAE performance on the individual domains is presented with dashed horizontal and vertical lines. The results from the POE model are presented as a Pareto front. For each domain combination we show plots for the performance on the Recall@50 and NDCG@50. In all cases, the best result is in the upper right corner. 

First, we train the POE model by giving equal weight to both domains in the loss function ($\forall d \in \{1,...,D\} \: \lambda_d = 1$ in Equation~\ref{elbo-loss}). The Pareto fronts are shown in blue on Figure~\ref{pareto-fronts}. We can observe that in some cases like Pizza and Burgers, the model trained with equal weights is improving upon the VAE performance on both domains simultaneously. However, on other cases like NDCG@50 on Restaurants and Shopping, the POE model is underperforming. 
Then, we assign different weights to the domains and train the POE model.  
In some cases like Books and Kindle or Books and Movies our model results in improved single-domain performance compared to the baseline model. However, on other cases 
increasing the weight of the Restaurant domain leads to a decrease in performance on that domain. 

The results show that in most cases, a set of optimal weights exists and we can improve upon the performance of VAE on a single domain of interest. Additionally, for a few combinations, 
we can improve simultaneously the performance on all domains. 
Finally, on some pairs, the POE model performs better in one domain only on Recall@50.

\begin{figure*}[t]
\centering
\begin{subfigure}[]{0.32\textwidth}
    \centering
    \includegraphics[width=\textwidth,height=0.9in]{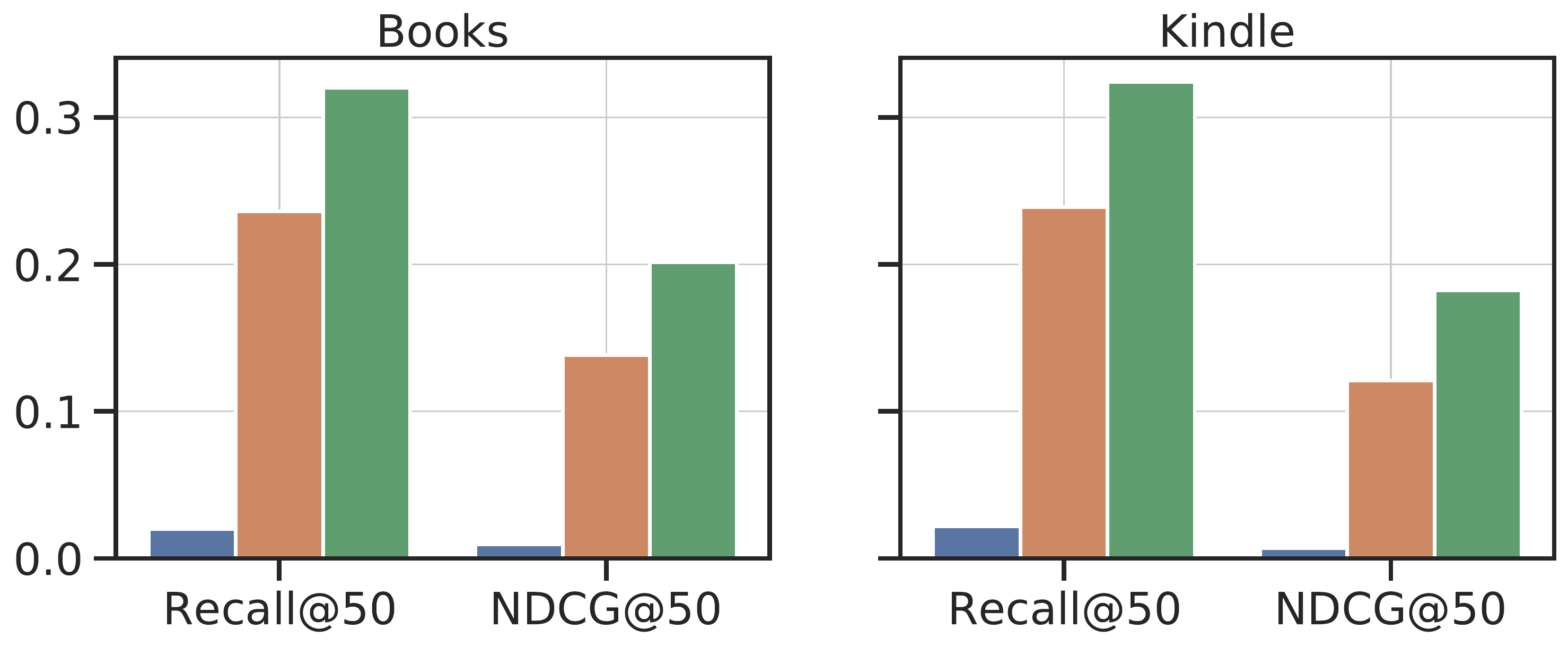}
    \caption{\textbf{Books and Kindle}.}
\end{subfigure}
~
\begin{subfigure}[]{0.32\textwidth}
    \centering
    \includegraphics[width=\textwidth,height=0.9in]{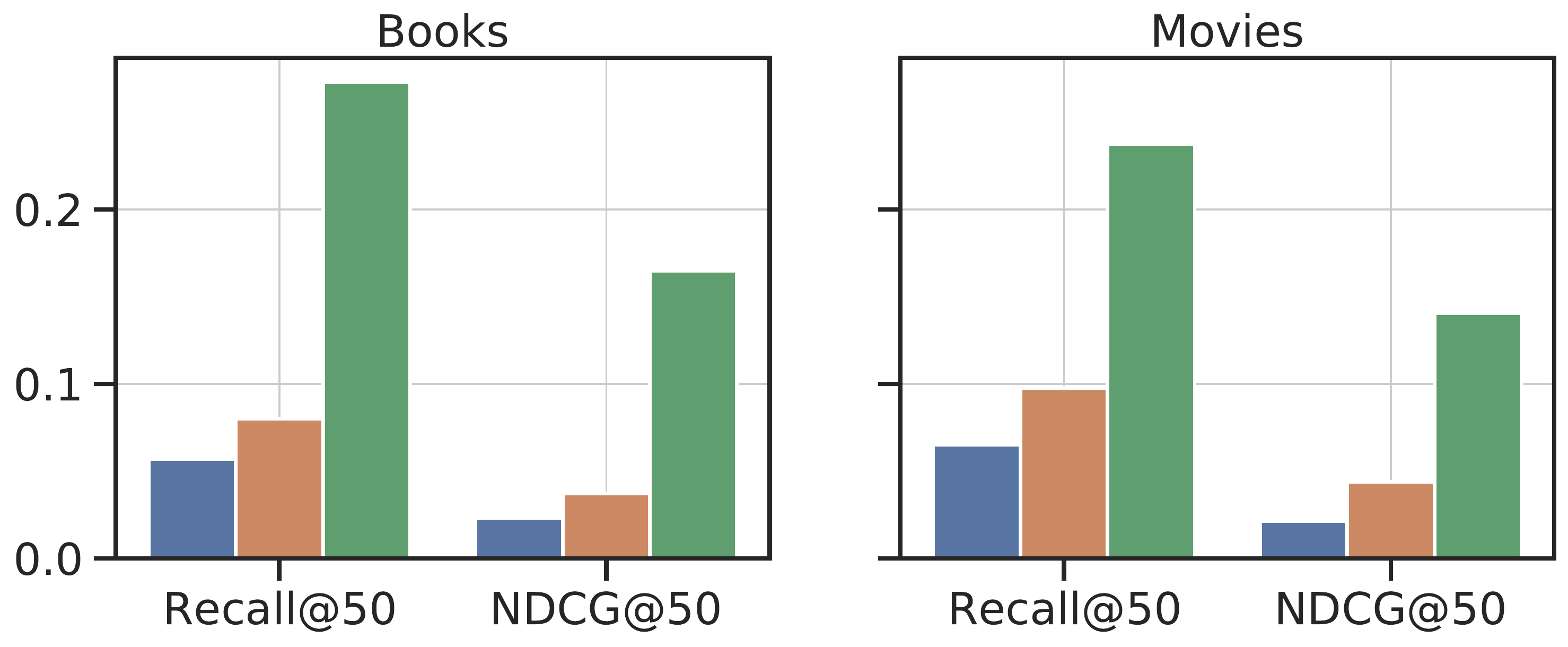}
    \caption{\textbf{Books and Movies}.}
\end{subfigure}
~
\begin{subfigure}[]{0.32\textwidth}
    \centering
    \includegraphics[width=\textwidth,height=0.9in]{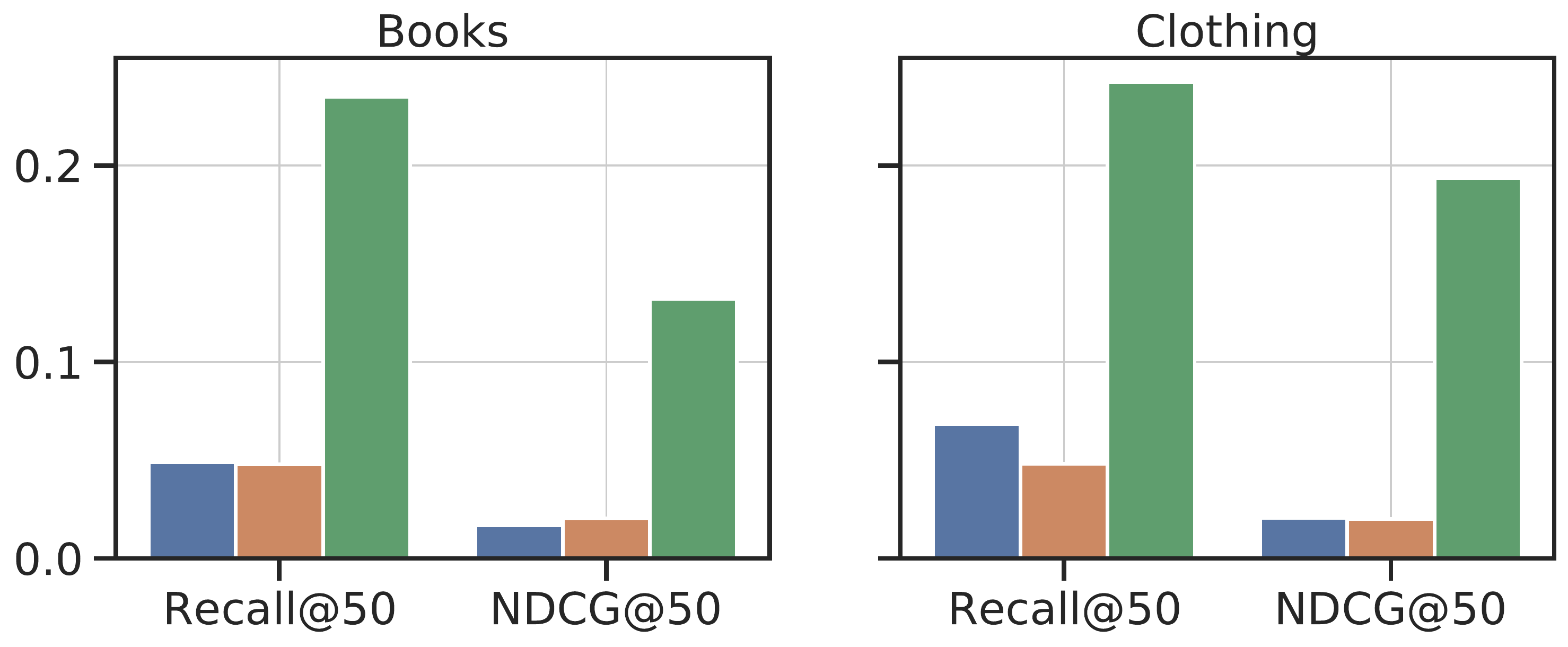}
    \caption{\textbf{Books and Clothing}.}
\end{subfigure}
\\
\begin{subfigure}[]{0.32\textwidth}
    \centering
    \includegraphics[width=\textwidth,height=0.9in]{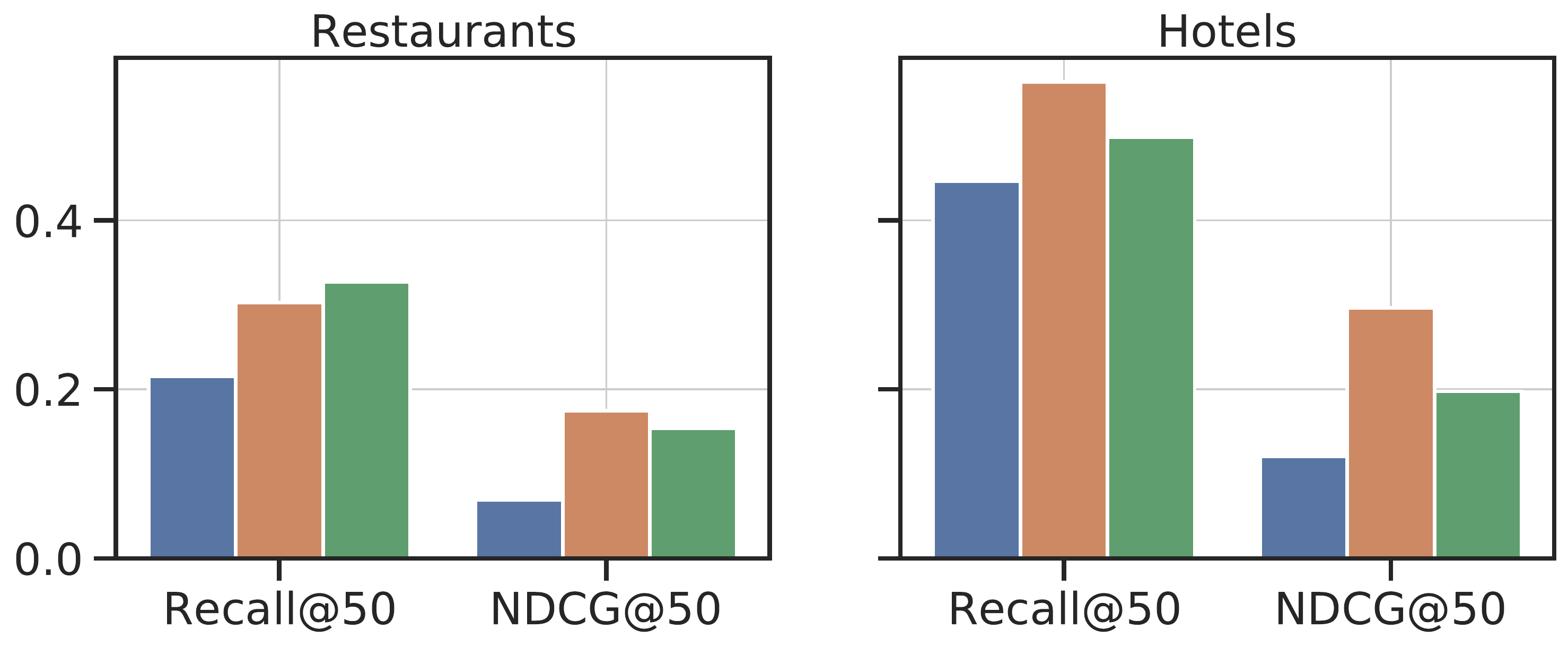}
    \caption{\textbf{Restaurants and Hotels}.}
\end{subfigure}
~
\begin{subfigure}[]{0.32\textwidth}
    \centering
    \includegraphics[width=\textwidth,height=0.9in]{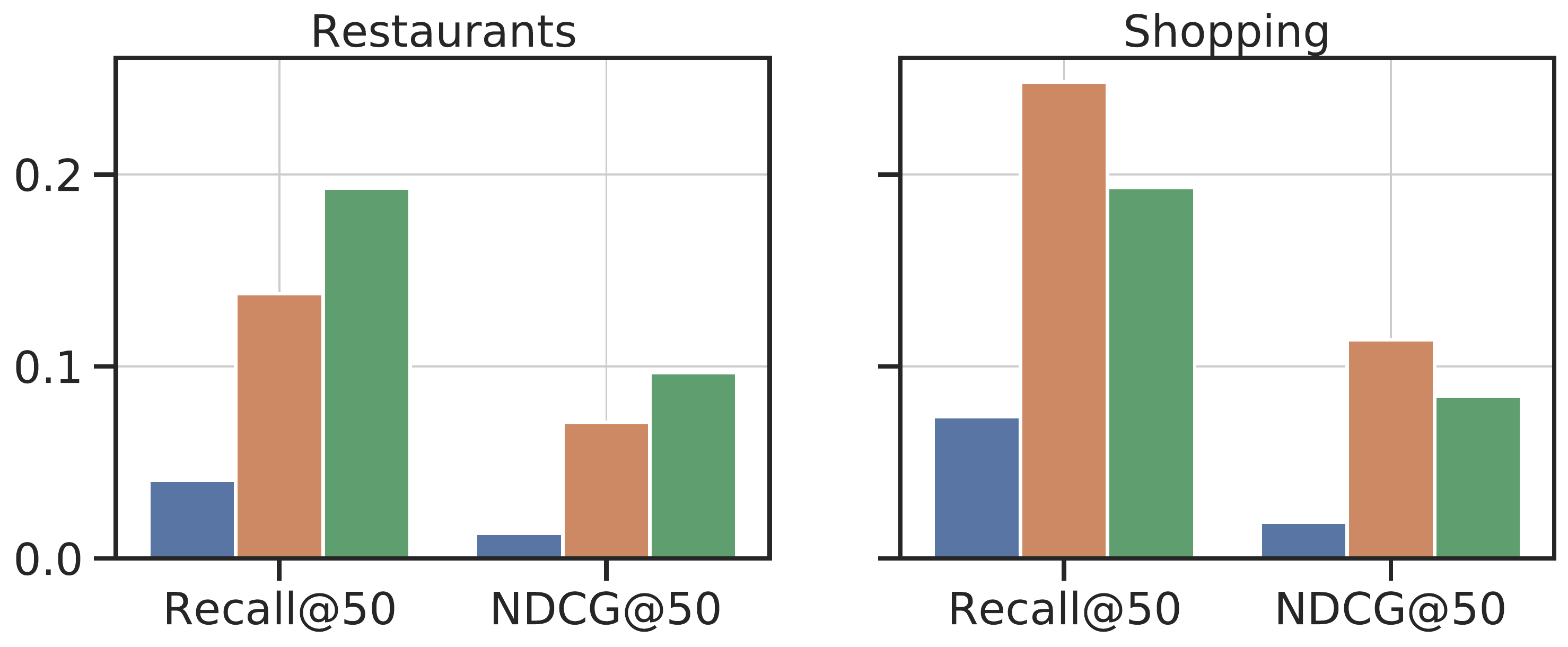}
    \caption{\textbf{Restaurants and Shopping}.}
\end{subfigure}
~
\begin{subfigure}[]{0.32\textwidth}
    \centering
    \includegraphics[width=\textwidth,height=0.9in]{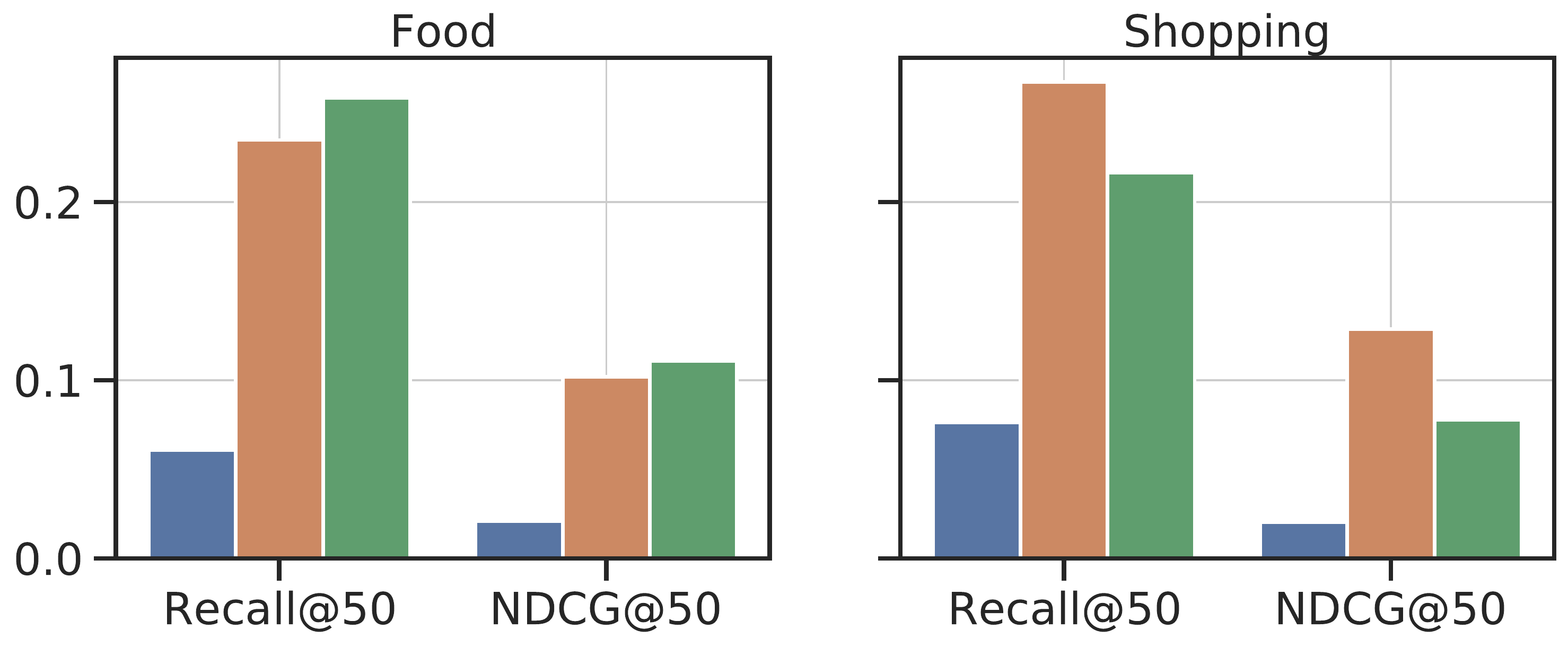}
    \caption{\textbf{Food and Shopping}.}
\end{subfigure}\\
\begin{subfigure}[]{0.32\textwidth}
    \centering
    \includegraphics[width=\textwidth,height=0.9in]{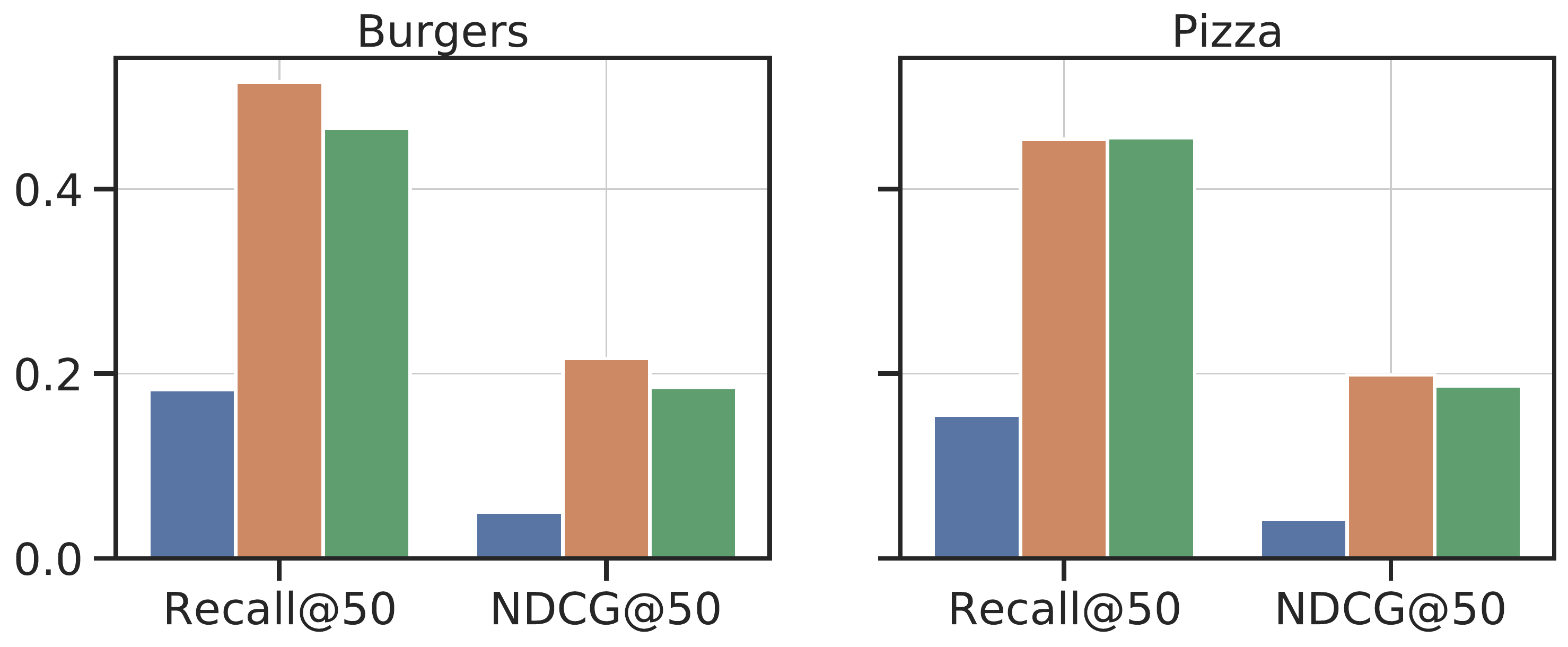}
    \caption{\textbf{Burgers and Pizza}.}
\end{subfigure}
~
\begin{subfigure}[t]{0.32\textwidth}
\end{subfigure}
~
\begin{subfigure}[t]{0.32\textwidth}
    \centering
    \includegraphics[width=0.5\textwidth]{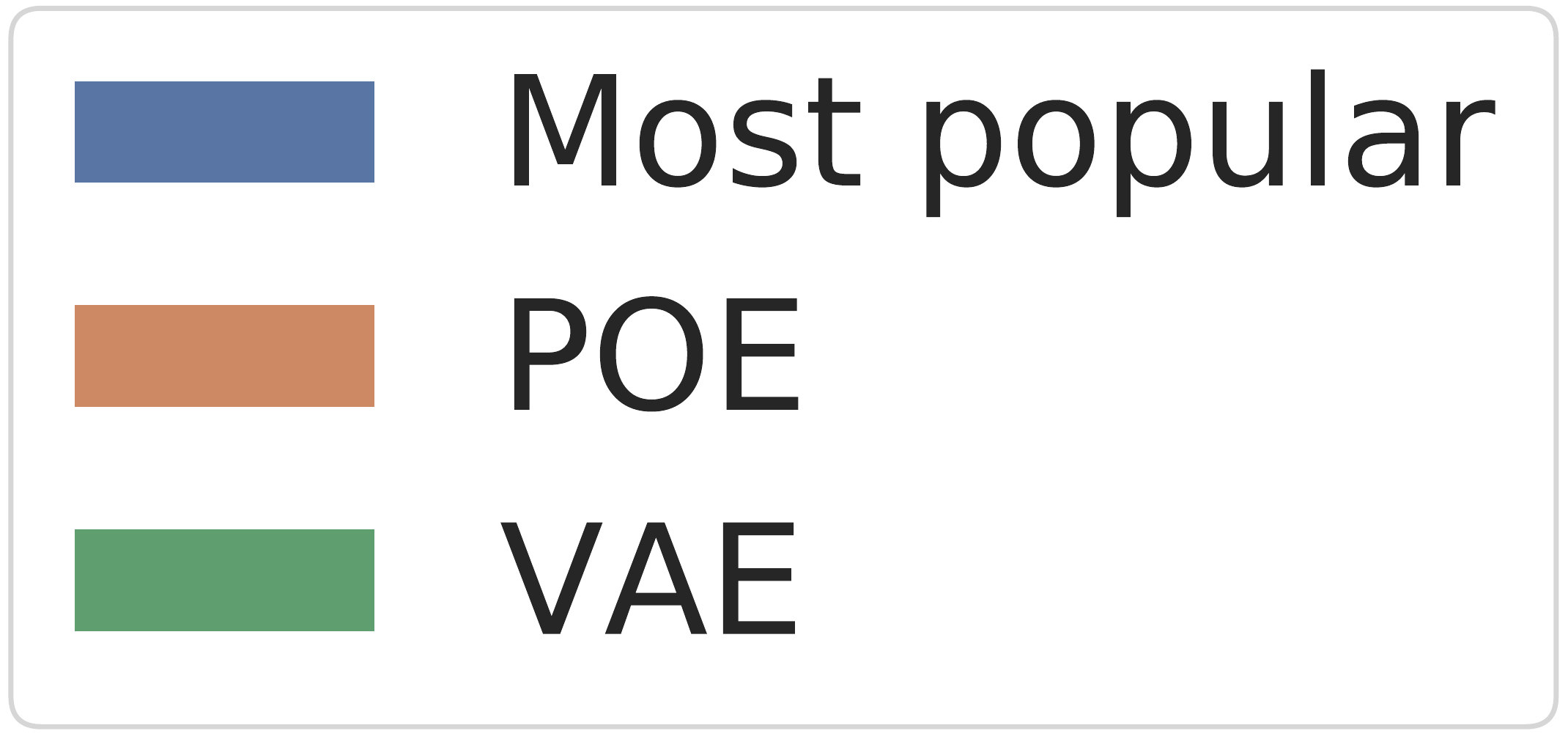}
\end{subfigure}
\caption{Performance of the POE model on the domain-specific cold-start problem.}
\label{pics:cross-domain-results}
\end{figure*}

\subsection{Cross-Domain Recommendation}

The cross-domain recommendation task is concerned with generating recommendation on the target domain by exploiting knowledge from the source domain. In our case, we exploit the ability of the POE model to generate recommendations for users with missing domains in order to tackle the cross-domain task. We perform the training and evaluation of POE on the intersection of users present in both domains. During evaluation, the model is shown user history from the source domain but evaluated on the target domain. In this way, we are able to simulate users with a missing domain and still evaluate the model by using users with known history on both domains. \\
\indent We compare the POE model with a baseline that recommends the most popular items in the dataset. Even though this is a very simple baseline, it is still widely used in practice in cases when there is no known history for a new user on the domain of interest. We also compare the POE model's performance with a single-domain VAE trained and evaluated on the target domain. Unlike the POE, the VAE is shown the input from the target domain during evaluation. In this way, we evaluate to which extend can the POE model transfer knowledge from one domain to the other. 

In Figure~\ref{pics:cross-domain-results}, we see that the POE model outperforms the \textit{Popular} baseline in most cases. 
On 4 out of 7 combinations the POE model is improving upon the VAE baseline on at least one domain. This means that the POE model can improve upon the single-domain recommendation performance of a standard VAE without seeing the user history on the domain of interest. 
We can conclude that the POE model is a viable solution to the domain-specific cold-start problem. 

\section{Conclusion}

We presented a method to alleviate data sparsity in creating recommendations. We base our approach on the single-domain variational autoencoder collaborative filtering and expand it to a multi-domain setup. We created a novel Product-of-Experts approach for joint preference modelling. We show that, using it, a user's preferences in a source domain can be used to improve recommendations in a target domain in two ways:  \begin{enumerate*}
    \item by complementing a shallow representation of the user in the target domain and
    \item in a cross-domain setup, where the recommendations are made solely based on the knowledge in the source domain.
\end{enumerate*}
The results point to an outperformance of POE over single-domain VAE approaches. Moreover, the cross-domain recommendations are competitive, even surpassing the in-domain recommendations in some cases.

\bibliographystyle{ACM-Reference-Format}
\bibliography{acmart}
\clearpage


\end{document}